\def\year{2020}\relax
\documentclass[letterpaper]{article} 
\usepackage{aaai20}  
\usepackage{times}  
\usepackage{helvet} 
\usepackage{courier}  
\usepackage[hyphens]{url}  
\usepackage{graphicx} 
\usepackage{graphicx}  
\title{
Riiid! Answer Correctness Prediction Kaggle Challenge:
\\
4th Place Solution Summary }
\author{
  	Duc Kinh Le Tran
\\
    Axance Technology, France
\\
    duckinh.letran@gmail.com
}

\begin{document}

\maketitle

\begin{abstract}
This paper presents my solution to the challenge "Riiid! Answer Correctness Prediction" on Kaggle hosted by Riiid Labs\footnote{\url{https://www.kaggle.com/c/riiid-test-answer-prediction/}}, which scores 0.817 (AUC) and ranks 4th on the final private leaderboard\footnote{Code to reproduce the solution is available at \url{https://github.com/dkletran/riiid-challenge-4th-place}}. It is a single transformer-based model heavily inspired from previous works such as SAKT, SAINT and SAINT+. Novel ingredients that I believed to have made a difference are the time-aware attention mechanism, the concatenation of the embeddings of the input sequences and the embedding of continuous features.
\end{abstract}

\section{Introduction}
Knowledge Tracing has been a very interesting problem from both the application and the academic point of view. The objective of Knowledge Tracing is to keep track of a student’s \emph{knowledge state} on an online learning platform \cite{pandey2019selfattentive}. The problem of Knowledge Tracing can be formulated like this: giving the history of learning activities of a student, predict if he/she answers correctly the question he/she is facing. This problem has been tackled by many methods from collaborative filtering to deep learning with recurrent networks.  Recently, transformers \cite{vaswani2017attention}, originally designed for language understanding, have been proven to outperform other aproaches. Some of these transformer-based models are SAKT \cite{pandey2019selfattentive}, SAINT \cite{choi2020appropriate} and SAINT+ \cite{shin2020saint}).

In this competition \cite{KaggleC}, with data the host made available \cite{choi2020ednet},  we have for each student the list of his/her questions in the past, along with his/her answers, the relative timestamp, time spent for each question and whether he/she viewed the explanation after answering the question. We also have a list of lectures  watched by users as they progress in their education. The metadata of questions and lectures can also be helpful in predicting the correctness of students' answers.

Like many other competitors I decided to use transformers for this challenge believing there is still room for improvement in existing transformer-based models. The other reason for using transformers in this competition is that the input data are much richer that those used in previous works. Effective integration of all potentially useful data on the inputs of the transformer's encoders and decoders could be the key to boost the performance. 

\section{Data Preparation}
To feed the transformer, raw train data in the form one activity per row,  need to be preprocessed. The first steps is to encode categorical features into integer indices for embedding layers of the transformer. This step is also applied to the metadata tables (question metadata and lecture metadata). On the train table, I also added \emph{time lag} - time interval from the last question bundle - which has been proven useful in the SAINT+ \cite{shin2020saint} (not exactly the same version of \emph{time lag} here but still very similar). Since all question of the same bundle (same container) share a timestamp, they also share one \emph{time lag}. 

Two statistics on the questions - \emph{question popularity} and \emph{question difficulty} - are also computed from the whole train table. Adding these two features was proven (experimentally) to be helpful, at least in terms of convergence speed of the training process. 

The train data table is then grouped by \emph{user id} into one row per student so that we have sequences representing the history of activities for each student. These sequences will be the inputs of the transformers' embedding layers.
 
\section{Model Architecture}

\begin{figure}[t]
\centering
\includegraphics[width=\columnwidth]{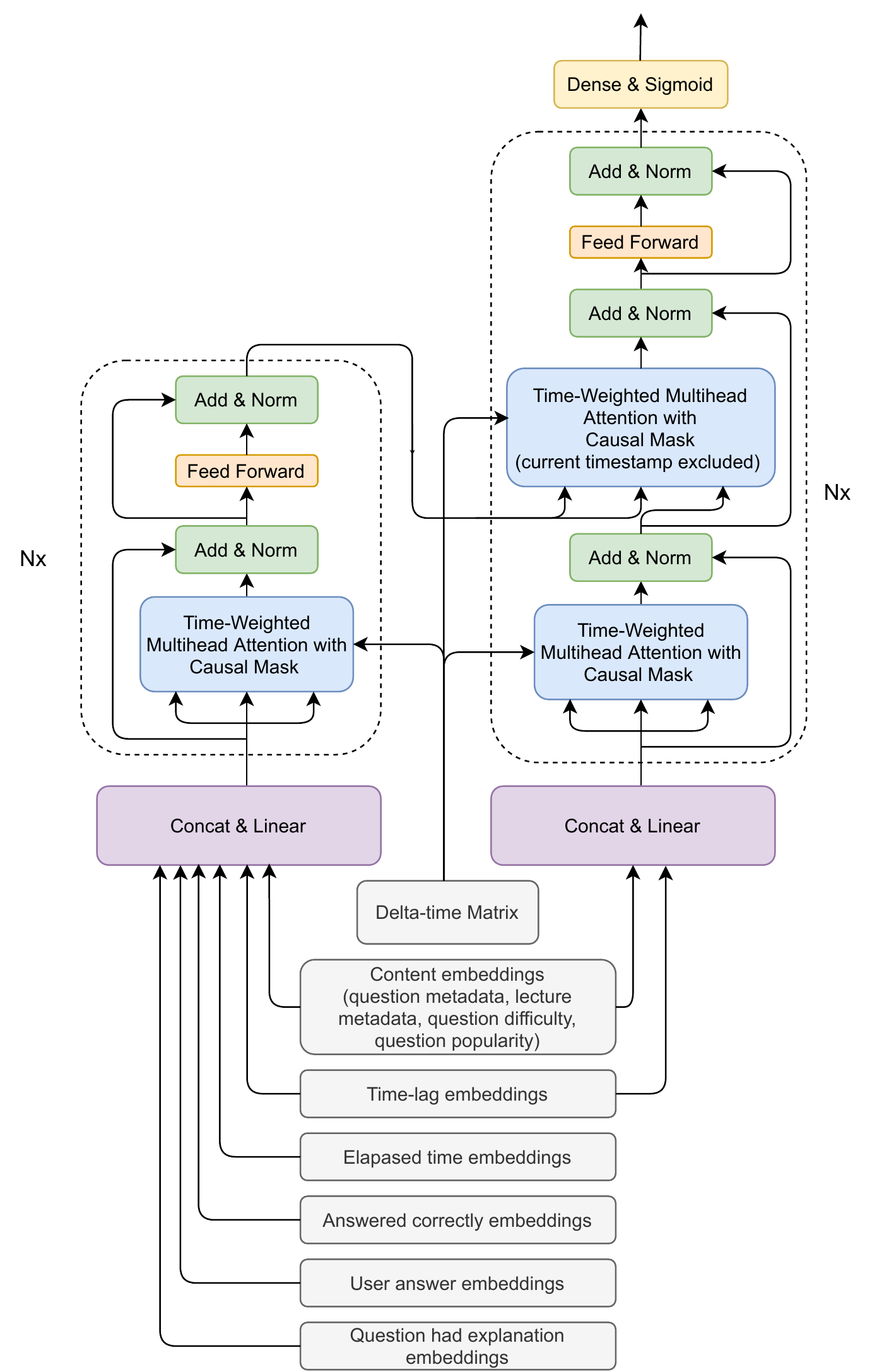}
\caption{Model architecture}.
\label{fig1}
\end{figure}

As shown in Figure \ref{fig1}, the model is designed to take into account all available input data that have been prepared from the previous step. Below is a walkthrough of the model architecture with details on its key features.
\subsection{Embedding Layers}
Each input sequence goes through an embedding layer with the same \emph{embedding size}.

Embeddings of the categorical features (\emph{answer correctly}, \emph{user answer}, \emph{question had explanation}) are normal embedding which is basically a lookup on the weight matrix by the indices in of the input sequences. 

Embedding of the content (questions and lectures) is computed using the \emph{ContentEmbeddingLayer}. The layer aims to encode all metadata of questions and lectures into the output embedding. Each element of the metadata of questions and lectures (for example tags, part, type) has a weight matrix and the embedding weight matrix of the layer is the concatenation of theses weight matrices. Two computed statistics of questions (\emph{question popularity} and \emph{question difficulty}) also contribute to the layer weight matrix. The output embedding is the result of the lookup on this concatenated matrix by the content index. 

\begin{figure}[t]
\centering
\includegraphics[width=\columnwidth]{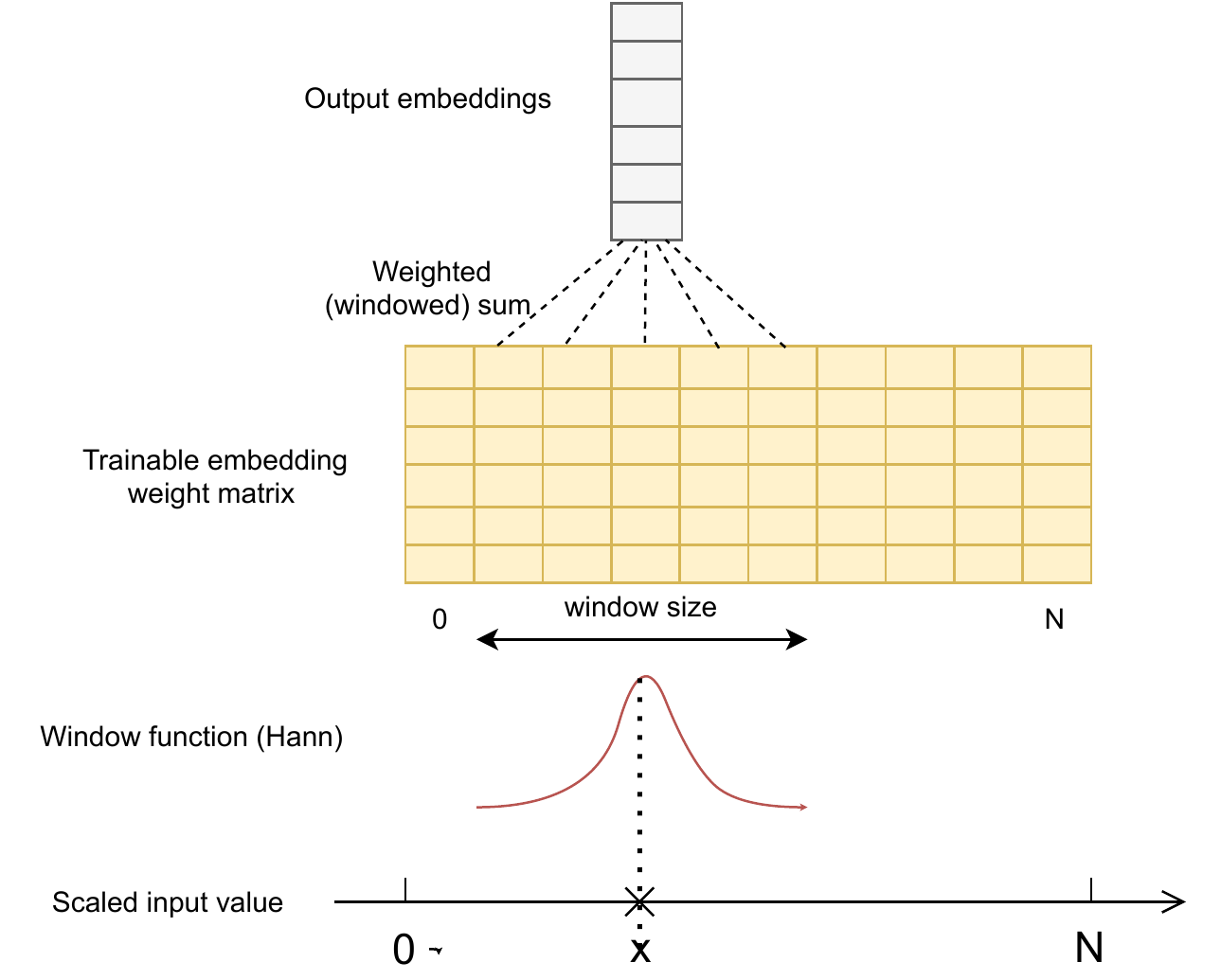}
\caption{Embeddings of continuous features}.
\label{fig2}
\end{figure}

Embeddings of continuous features (\emph{time lag, question elapsed time, question difficulty, question popularity}) are computed using a \emph{ContinuousEmbedding} layer. The computation of this layer is illustrated Figure \ref{fig2}. The main idea is to sum up (with weighted windowed sum) a number of consecutive embedding vectors of  the embedding weight matrix. The \emph{window size} is a hyper-parameter of the layer. The output of the layer is a "smooth" embedding vector of the continuous feature, in the sense that 2 values very close together should have similar embeddings.  In the experimentation, I observed that this method of embedding gave the model a little edge in performance in comparison with traditional \emph{categorical embedding} (where continuous features are descretized then gone through normal a embedding layer). 
\subsection{Encoder and Decoder Input}
Embeddings of all input sequences are concatenated and linear-transformed to feed the first layer of the encoder and that of the decoder of the transformer. As shown in the schema, input of the first encoder layer contains embeddings of all input elements and input of the decoder doesn't not contain user answer related elements (only \emph{content embeddings} and \emph{time lag embeddings}).  This is an intuitive choice justified by the following:
\begin{itemize}
\item The encoder input value of a position represents the knowledge state of the student up to that position. We need all information (content, time lag, his/her answer, answer correctness, time spent on the question and whether he/she watched the explanation after answer the question).
\item The decoder input (query) at a position represents the exercise (question) and available context information at that position. We only have information about the content itself and \emph{time lag} from the previous exercises.
\item To predict the user answer (its correctness) at a position we need to know his/her "knowledge" and the current question and context. That is why the both the encoder output and the decoder query are combined in the decoder layer to form the output of the model.
\item Attention and self-attention of the encoder and decoder layers help to look back and take into account not not only the most recent events but the whole history of students' activities.
\end{itemize}

\subsection{Attention And Self-Attention Mechanism}
Encoder and decoder layers are almost identical to those in the transformer original paper \cite{vaswani2017attention}. The differences reside in the multi-head attention (and self-attention) part. Fisrt of all, to prevent the current position from seeing future inputs, we must apply causal masks to the self-attention of encoder and decoder layers. This causal mask is computed from the timestamp sequence: a position at timestamp $t_0$ only attends to positions with timestamp $t \ge t_0$ (which means it can attend to itself and other interactions of the same container). Attention mask of the encoder output to the decoder is a little different because the decoder at a position cannot see user answer related inputs at that position. This means that the encoder output at a timestamp $t_0$ can only attend to decoder positions that are strictly in the future $t > t_0$ (i.e not including the current timestamp).

The other feature that gives the model some boost in performance is the \emph{Time-Weighted Multihead Attention}, an improvement of the original \emph{Multihead Attention} based on the observation that events in a distant past should attend less to the current position than recent events do. The idea is to decay the attention coefficients by a factor of $T_{ij}^{-w}$ where $T_{ij} = t_{j} - t_{i}$ is the difference in timestamp of a position $i$ and the position $j$ it attends to and $w$ is a non-negative trainable parameter. This is implemented by modifying the "Scaled Dot-Product Attention" of the encoders and decoders. In the original paper this "Scaled Dot-Product Attention"  is:
\begin{equation}
\mathrm{Attention}(Q,K,V) = \mathrm{softmax}( {QK^T \over \sqrt{d_k}})V
\end{equation}

Since the dot product $QK^T \over \sqrt{d_k}$ is in logarithmic scale with respect to the attention coefficients (the softmax of the dot prodcut), decaying the attention coefficients by a factor of $T_{ij}^{-w}$ is equivalent to substracting $w  \log {T_{ij}}$ from the dot product. In other words, the "Scaled Dot-Product Attention" can be rewritten as the following to take into account the time decay factor:

\begin{equation}
\mathrm{Attention}(Q,K,V) = \mathrm{softmax}( {QK^T \over \sqrt{d_k}} - w\log T)V
\end{equation}

where $T$ is the timestamp difference matrix  ($T_{ij} = t_{j} - t_{i}$ with $t_i, t_j$ is the timestamp at the position $i$ and $j$ respectively). In practice, $\log T$ is computed once and then used as an input to all the encoder and decoder layers.

As stated earlier, $w$ is a trainable parameter constrained to be non-negative. We have one parameter per attention head.  When tuning the model, I found it works best when $w$ is randomly initialized in the interval (0,1).
\section{Training}
The output of the transformer goes through a dense layer with \emph{sigmoid} activation after which we get the predicted probability of the correctness of the user answer to the question at each position. \emph{Binary Cross-Entropy Loss} is used to train the model parameters. Positions where the content is a lecture are masked out to be excluded from the loss function. 

Input sequences are cut and padded to have the same length.  To put it more concretely, short sequences are padded to have the required length and long sequences are randomly cut into smaller parts, each of which is kept and padded if necessary to the required length. In training, the sequence length is set to 1024.

The model was implemented in Tensorflow and trained on  Colab Pro TPU. To monitor the model performance locally and also for hyper-parameter tuning, I used the validation strategy by \citeauthor{riiidcv}  \cite{riiidcv}.  This is basically a train-validation split which leaves 2.5\% of the train set (around 2.5M iteractions) for validation. It is a very reliable validation strategy thanks to the good quality of the data. Throughout the competition, all improvements measured on the validation set lead to improvement on the leader board.

The final version of the model has embedding size of 128, model size of 512, 4 encoder layers and 4 decoder layers. Adam optimizer is used with $\beta_1=0.9, \beta_2=0.98, \epsilon=1e-9$. Batch size is set to 64 and dropout rate is 0.0. Learning rate, initially set to $2.5 e -4 $ is scheduled to have a warm up stage of 4000 steps followed by cosine decay stage of around 30000 steps. Training took around 4 hours. The AUC on the validation set is around 0.818 and the accuracy is 75.35\% at the end of training.

The model used for my final submission (scored 0.817 AUC) was trained on the whole train dataset with the same  configuration as above. 

\section{Inference}
Due to resource limitation (9 hours of execution with GPU to make prediction for 2.5M rows), in the inference kernel, the sequence length is reduced to 512. Prediction in inference kernel, constrained by the time-series API, is not entirely in the same condition as in training and validation : we only predict the correctness of students' answers at the latest timestamp (the last batch of questions). For each student, history in a window of 512 latest activities are kept in the memory (in form of numpy arrays). In  each iteraction, these activity histories are updated and fed into the model to predict the correctness of the student's answer to the last question (or a few questions in the last batch). Inference on the whole test set took around 8.5 hours, estimated from the moment the kernel get submitted and the moment its score is available on the leaderboard.

\section{Observation and Discussion}
The most challenging part of training the transformer is to avoid overfitting, which comes after 1-2 epochs. Various methods like dropout, label smoothing, partial label masking have been tried out but all led to no obvious improvement or worse results.  

It turns out that reducing the embedding size and concatenating the embeddings of different input elements work much better than having the embedding size equal to the model dimension and  summing the embeddings. Reducing  the embedding size obviously has some kind of  regularization effect. Also, in the final version of the model, I added a \emph{layer normalization} layer \cite{ba2016layer} after the content embedding layer. It helps regularize the layer weights and give a small improvement in performance. 

Two other features proposed in the model (\emph{Time-Weighted Atttention} and embedding of continuous features also contribute to improve AUC. Without the \emph{Time-Weighted Atttention} the model scores 0.816 AUC on the validation set and without the proposed embedding layer for continuous features (using categorical embedding instead), the model's AUC is 0.817 on the validation set. As mentioned earlier,  the model's AUC on the validation set is around 0.818 when both of the proposed features are activated. This is not a big improvement but still significant in the context of this competition.

Longer sequence length also helps. Due to resource limitation I ended up training the model with the size of 1024 and reduce it to 512 in the submission kernel. What is interesting is that training with sequence length 1024 then doing inference with 512 works  better than training  with 512 and then doing inference with 512 (same length) (the gap is around 0.002 AUC). 

My choice of model size (model dimension 512 and and number of layers 4) was influenced by previous work (SAINT, SAINT+), but it turns out this is a good choice in terms of compromising the performance and computational resources. With this configuration, the model has just enough time (within 9h limitation) to do inference on the entire test set (with the sequence length of 512) in the submission kernel so there's no need to increase it further. 

\bibliographystyle{aaai} \bibliography{Model_Description}

\end{document}